\documentstyle[aps]{revtex}
\begin{document}
\draft
\title{Effective potential for the order parameter of the SU(2)
Yang-Mills deconfinement transition}
\author{Michael Engelhardt
and Hugo Reinhardt }
\address{Institut f\"ur theoretische Physik \\
Universit\"at T\"ubingen \\
Auf der Morgenstelle 14 \\ 72076 T\"ubingen, Germany }
\date{ }
\maketitle

\begin{abstract}
The Polyakov loop variable serves as an order parameter to characterize
the confined and deconfined phases of Yang-Mills theory. By integrating
out the vector fields in the SU(2) Yang-Mills partition function in
one-loop approximation, an effective action is obtained for the
Polyakov loop to second order in a derivative expansion. The resulting
effective potential for the Polyakov loop is capable of describing a
second-order deconfinement transition as a function of temperature.
\end{abstract}

\pacs{PACS: 11.10.Wx, 12.38.Aw, 12.38.Lg}

{\bf Introduction.}
Confinement of quarks and gluons into color-singlet clusters is the most
important characteristic of strong interaction physics. The phenomenon
has been observed in calculations based on the lattice formulation 
\cite{mm} of Yang-Mills theory, and diverse physical mechanisms 
for the emergence of a confining potential between color sources have 
been proposed, among others, 
the dual Meissner effect picture of 't Hooft and Mandelstam \cite{mag},
random fluxes \cite{olesen}, the stochastic vacuum \cite{dosch},
the leading-log model of Adler \cite{adler}, and dual QCD \cite{zacha}.

On the other hand, if one generalizes to finite temperature, it is
believed (and borne out e.g. by lattice calculations \cite{mm}) that
there exists a transition to a phase in which color sources become
deconfined. An order parameter distinguishing the two phases in
pure gauge theory is the Polyakov loop \cite{polya}, \cite{suss}
\begin{equation}
L(x) = \frac{1}{N_C } \mbox{tr T} \exp \left(
i\int_{0}^{\beta } dx_0 \, A_0 (x_0 , x) \right) \ ,
\label{poly}
\end{equation}
where $\beta $ is the extension in time direction of the (Euclidean)
space-time manifold under consideration. It can be identified with
the inverse temperature of the ensemble in which the expectation
value of $L$ is taken. $A_0 $ denotes the time component of the
gauge field, $T$ time ordering, and $N_C $ the number of colors.
The negative logarithm of the expectation value of the
Polyakov loop can be interpreted \cite{ben} as the free energy
associated with a single static color source in the fundamental
representation of the gauge group. Vanishing of $\langle L\rangle $
indicates infinite free energy, i.e. confinement; finite 
$\langle L\rangle $ indicates deconfinement. There is a loophole
to the former interpretation: The free energy may be infinite
due to reasons not connected with the infrared phenomenon of confinement,
e.g. an infinite self-energy, which is an ultraviolet phenomenon.
One should thus take care to separate infrared and ultraviolet
effects in the evaluation of $\langle L\rangle $.
The behavior of $\langle L\rangle $ is intimately connected to a
discrete symmetry of Yang-Mills theory called center symmetry
\cite{ben}. This symmetry leaves the Yang-Mills action invariant but
changes $L$ by a phase $e^{2\pi i/N_C } $.
Therefore, if the symmetry is realized,
one automatically has $\langle L\rangle =0$, i.e. confinement; if the
symmetry is broken, one has $\langle L\rangle \neq 0$, deconfinement.
Which scenario is realized can be read off by calculating the effective
action governing the Polyakov loop order parameter and interpreting
it in the spirit of Ginzburg-Landau theory \cite{ben}. Such
considerations have been quite successful in the lattice
formulation \cite{polon}-\cite{kurt}, where, however, as always
the continuum limit represents a problem. A continuum calculation
was carried out in \cite{weiss}, which was only able
to capture the deconfined phase (see however also \cite{meis}, in which
a model built on ideas familiar from the Savvidy vacuum \cite{savi}
is suggested).

In this letter, a continuum evaluation of the effective action
is exhibited for the case of SU(2) color which goes significantly
beyond previous work and in contradistinction to \cite{weiss}
describes both the confined and deconfined phases.
Specifically, centrifugal barriers in kinetic terms which were not
considered in \cite{weiss} substantially modify the effective
potential for the Polyakov loop order parameter. The success of
this calculation gives rise to the hope that the confined phase can 
be understood essentially from the metric properties of the gauge
group without extensive recourse to details of the dynamics.

{\bf Approximation of the SU(2) effective action.}
A particularly convenient gauge for discussing the behavior
of the Polyakov loop variable (\ref{poly}) in the case of
SU(2) color is \cite{weiss}, \cite{hugo}-\cite{lell}
\begin{equation}
\partial_{0} A_0 = 0 \ \ \mbox{and} \ \ 
A_0 = \mbox{diag} \, (a_0 /2 , -a_0 /2 ) \ .
\end{equation}
In this gauge, the Polyakov loop takes the simple form
$L=\cos \beta a_0 /2$ and center symmetry relates $a_0 $ and
$2\pi /\beta -a_0 $ if $a_0 \in [0,2\pi /\beta ]$. Thus, in order to gain
insight into the dynamics of the Polyakov loop, one should integrate
out in some approximative scheme the vector fields $A_i^a $ in the
SU(2) Yang-Mills partition function \cite{hugo}, \cite{lell}
\begin{equation}
Z= \int [Da_0 ] \det \mu (a_0 ) \int [DA_i^a ] e^{-S_{YM} [a_0 , A_i^a ] }
\end{equation}
(without much loss of generality, only the case of vanishing vacuum angle
$\Theta $ is considered here). The measure $\det \mu (a_0 ) $ denotes
the Cartan part of the SU(2) Haar measure for $a_0 $ at every space point.

The approximation to be used here is the following: Terms in $S_{YM} $
of higher order than quadratic in the vector fields are dropped,
and the resulting functional determinant is
calculated in a gradient expansion up to terms quadratic in derivatives
of $a_0 $. Previous work \cite{weiss} only considered constant $a_0 $.
It will become apparent that centrifugal barriers in the kinetic terms 
lead to a substantial modification of the effective potential.
The effective action for $a_0 $ in the above approximation can be 
written as
\begin{equation}
S_{eff} [a_0 ] = \frac{\beta }{2g^2 } \int d^3 x \, \partial_{i} a_0
\partial_{i} a_0 - \ln \det \mu (a_0 ) - \ln \mbox{det}^{-1/2} \Delta \ ,
\label{effac}
\end{equation}
where the last term stems from the Gaussian integration over the vector
fields $A_i^a $. The operator $\Delta $ becomes block-diagonal in color
space if one changes to the variables $(A_i^1 \pm iA_i^2 )/\sqrt{2} $,
$A_i^3 $. Then one has two blocks
\begin{equation}
\Delta^{\pm }_{ij} = -\delta_{ij} \partial^{2} + \partial_{i} \partial_{j}
+\delta_{ij} (-i\partial_{0} \pm a_0 )^2 \ ,
\label{deldef}
\end{equation}
and a third block independent of $a_0 $ which gives only an irrelevant
constant.

{\bf Evaluation of the determinant.}
Concentrating on the operator $\Delta^{-} $, its functional determinant
can be expressed in the form \cite{ramond}
\begin{equation}
\ln \mbox{det}^{-1/2} \Delta^{-} = 
\frac{1}{2} \left. \frac{\partial }{\partial s}
\frac{1}{\Gamma (s)} \int_{0}^{\infty } dt \, t^{s-1} \beta \int d^3 x \,
\mbox{tr} K(x_0 ,x;x_0 ,x;t) \right|_{s=0}
\label{zeta}
\end{equation}
with the heat kernel
\begin{equation}
K_{ij} (x_0 ,x;y_0 ,y;t) = \left( e^{-t\Delta^{-} } \right)_{ij}
\delta^{3} (x-y) \delta (x_0 -y_0 ) \ ,
\label{heatdef}
\end{equation}
where $\Delta^{-} $ is evaluated at, and acts on, the coordinates
$x$ and $x_0 $. In the limit $(x-y)\rightarrow 0 $, one can expand
\begin{equation}
a_0 (x) = a_0 (y) + (x-y)_{i} \partial_{i} a_0 (y) + \frac{1}{2}
(x-y)_{i} (x-y)_{j} \partial_{i} \partial_{j} a_0 (y) \ .
\end{equation}
Higher orders in derivatives of $a_0 $ are neglected, as mentioned above.
Upon inserting Fourier representations for the $\delta $-functions
in (\ref{heatdef}) and commuting $\exp (-t\Delta^{-} )$ through to 
the right, one obtains the representation
\begin{equation}
K_{ij} (x_0 ,x;y_0 ,y;t) = 
\frac{1}{\beta (2\pi)^3 } \sum_{r=-\infty }^{\infty } 
\int d^3 k \, e^{i(k(x-y) + 2\pi r (x_0 -y_0 )/\beta )}
e^{-tk^2  } e^{-t(2\pi r/\beta -a_0 (y) )^2 }
\left( e^{-(A+B)} \right)_{ij}
\end{equation}
with
\begin{eqnarray}
A_{ij} &=& -t k_i k_j \\
B_{ij} &=& t(-\delta_{ij} (2ik_l \partial_{l} +\partial^{2} )
+ik_i \partial_{j} +ik_j \partial_{i} +\partial_{i} \partial_{j}
+\delta_{ij} (x-y)_m C_m +\delta_{ij} (x-y)_m (x-y)_n D_{mn} ) \\
C_m &=& 2(a_0 (y) -2\pi r/\beta ) \partial_{m} a_0 (y)
\label{cdef} \\
D_{mn} &=& (a_0 (y) -2\pi r/\beta ) \partial_{m} \partial_{n} a_0 (y)
+\partial_{m} a_0 (y) \partial_{n} a_0 (y) \ .
\label{ddef}
\end{eqnarray}
$A$ and $B$ are to be interpreted as operators acting to the right on unity.
Now one can expand in powers of $B$, since only terms up to quadratic in
derivatives of $a_0 $ are to be retained. One needs terms up to fourth 
order in $B$. Since $A$ and $B$ do not commute, the expansion
\begin{eqnarray}
e^{-(A+B)} &=& e^{-A} - \int_{0}^{1} ds_1 \, e^{-(1-s_1 )A} B e^{-s_1 A} 
\label{expan} \\
& & +\int_{0}^{1} ds_2 \int_{0}^{1} ds_1 \, s_1 e^{-(1-s_1 )A} B
e^{-s_1 (1-s_2 )A} B e^{-s_1 s_2 A} - \ldots
\nonumber
\end{eqnarray}
must be used.
In the limit $x-y\rightarrow 0$, only terms survive in which all factors
$(x-y)_i $ are removed by derivative operators. 
The matrix algebra is carried out using
$ (\exp (-wA) )_{ij} = \delta_{ij} + (\exp (wtk^2 ) -1 ) k_i k_j /k^2 $.
Furthermore evaluating the integrals over the $s_i $ in (\ref{expan}),
one arrives at
\begin{eqnarray}
\mbox{tr} \, K (x_0 ,x;x_0 ,x;t) 
&=& \frac{1}{\beta (2\pi )^3 } \int d^3 k \,
\sum_{r=-\infty }^{\infty } e^{-t(2\pi r/\beta -a_0 )^2 } \left[
2e^{-tk^2 } +1 \right. \nonumber \\
& & +(2t(1+e^{-tk^2 } )/k^4 -4(1-e^{-tk^2 } )/k^6 ) \cdot
(k_i (D_{ij} - C_i C_j t/4)k_j -k^2 D_{ii} + k^2 C_i C_i t/4 ) \nonumber \\
& & \left. +e^{-tk^2 } (8k_i D_{ij} k_j t^3 /3 -k_i C_i C_j k_j t^4
-2D_{ii} t^2 + 2C_i C_i t^3 /3 ) \right] \ .
\end{eqnarray}
In order to regularize the momentum integrations in the
ultraviolet, one needs to introduce a cutoff $\Lambda $ on $|k|$.
Subsequently expanding the terms in the
square brackets in powers of $t$, one can do
the $t$-integration called for in (\ref{zeta})
using $\int_{0}^{\infty } dt \, e^{-ta} t^{s-1+n}
= \Gamma (s+n) / a^{s+n} $. After dividing by $\Gamma (s) $ and
taking the derivative w.r.t. $s$ at $s=0$, one can resum the expansions
(these are derivatives and integrals of geometrical series).
After explicitly inserting
$D_{ii} $ and $C_i C_i $ (cf. (\ref{ddef}) and (\ref{cdef})) in order
to make all Matsubara frequencies manifest, one can finally carry out the
Matsubara summations using
\begin{eqnarray}
\sum_{r} \frac{1}{(2\pi r/\beta +x)^2 +y^2 } &=& \frac{\beta }{2y}
\frac{\sinh \beta y}{\cosh \beta y -\cos \beta x } \ , \\
\sum_{r} \frac{1}{2\pi r/\beta -x} &=& -\frac{\beta }{2}
\cot (\beta x/2) \ ,
\end{eqnarray}
and appropriate derivatives and integrals over $x$ and $y$ thereof.
One finally arrives at
\begin{eqnarray}
\ln \mbox{det}^{-1/2} \Delta &=& 2 \cdot \frac{1}{2} \beta \int d^3 x \,
\frac{1}{\beta } \left[ -\frac{\Lambda^{3} }{6\pi^{2} } \ln 
\sin^{2} \pi c -\frac{1}{\pi^{2} \beta^{3} } \int_{0}^{\beta \Lambda }
dy \, y^2 \ln (\cosh y -\cos 2\pi c ) \right. \nonumber \\
& & +\frac{1}{2\beta } \partial_{i} c \partial_{i} c \cdot \left(
-\frac{4\beta \Lambda }{3\sin^{2} \pi c}
-\frac{13}{9} \frac{\sinh \beta \Lambda }{\cosh \beta \Lambda
-\cos 2\pi c}
+\frac{11}{3} \int_{0}^{\beta \Lambda } \frac{dy}{y}
\frac{\sinh y}{\cosh y -\cos 2\pi c} \right. \label{ergeb} \\
& & \left. \left. \ \ \ \ \ \ \ \ \ \ 
+\frac{4\beta \Lambda }{9} 
\frac{1-\cos 2\pi c \cosh \beta \Lambda }{(\cosh \beta \Lambda
-\cos 2\pi c)^2 }
-\frac{\beta^{2} \Lambda^{2} }{9}
\frac{\sinh \beta \Lambda (2-\cos^{2} 2\pi c
-\cosh \beta \Lambda \cos 2\pi c) }{(\cosh \beta \Lambda
-\cos 2\pi c)^3 } \right) \right] \ ,
\nonumber
\end{eqnarray}
where one derivative has been shoveled off terms $\partial_{i}
\partial_{i} a_0 $ by means of a partial integration, the rescaled
variable $c=\beta a_0 /2\pi $ has been introduced, and the factor 2
in front comes from adding the contribution of
the determinant of $\Delta^{+} $ (it turns out that
$\det \Delta^{-} $ and $\det \Delta^{+} $ separately are symmetric in 
$a_0 $). The first term in the square bracket in (\ref{ergeb}) exactly
cancels the term $-\ln \det \mu (a_0 )$ in the effective action 
(\ref{effac}) stemming from the Haar measure. This cancellation was 
already observed for space-time independent fields in \cite{weiss}. 
In that case, after splitting the vector fields into their transverse 
and longitudinal parts, one sees directly that the longitudinal 
subdeterminant cancels the Haar measure.

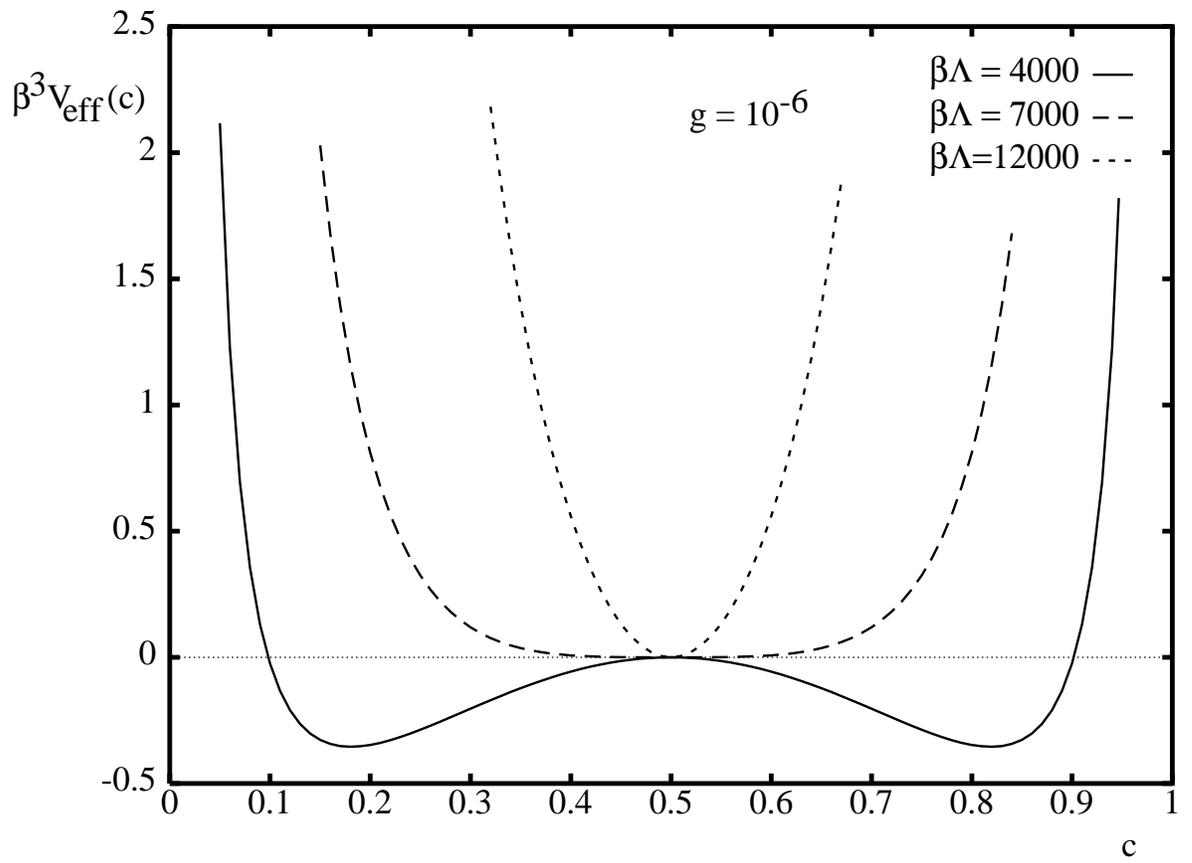
\begin{figure}
\caption{Effective potential, up to an irrelevant constant shift,
as a function of $c=\beta a_0 /2\pi$.}
\label{fig}
\end{figure}

{\bf Interpretation and discussion.}
The calculation sketched above has yielded an effective action of the
form
\begin{equation}
S_{eff} [c] = \int d^3 x \, 
\left( \frac{1}{\beta^{3} } V(c) + 
\frac{1}{\beta } W(c) \partial_{i} c \partial_{i} c \right)
\label{eform}
\end{equation}
for the field $c$. $V(c)$ is (up to a factor $-\beta^{3} $) the second term
in the square brackets in (\ref{ergeb}), and $W(c)$ is the term in the large
parentheses in (\ref{ergeb}) (up to a factor $-1/2$),
supplemented by a constant from the tree level contribution
in (\ref{effac}). It should be emphasized that the preferred mean
field value of $c$ cannot be read off from $V(c)$ alone. When evaluating
path integrals over $c$, only the class of space-time constant
configurations will give its dominant contribution at the value
minimizing $V(c)$. The weight of all other configurations is
influenced by $W(c)$, which may strongly penalize values preferred
by $V(c)$. In order to properly extract an effective 
potential from this action, one should transform the weight function
$W$ away from the kinetic term by the local change of variables
$u(c) = \int_{1/2}^{c} dc^{\prime } \, \sqrt{W(c^{\prime } )/\beta } $.
The action in the new variables $u$ then acquires the standard form
$\tilde{S}_{eff} [u] = \int d^3 x \, ( \partial_{i} u \partial_{i} u
+V_{eff} (c(u)) )$,
where
\begin{equation}
V_{eff} (c(u)) = \frac{1}{\beta^{3} } V(c(u)) 
+\frac{(\beta \Lambda )^3 }{12 \pi^{2} \beta^{3} }
\ln W(c(u))
\label{veffu}
\end{equation}
can now be straightforwardly identified as an effective potential. The
second term in (\ref{veffu}) comes from the Jacobian of the change of
variables in the path integral. Since one is ultimately interested in the
preferred value of $c$, one may minimize this potential directly as
a function of $c$. Note that in the $\beta \Lambda \rightarrow \infty $
limit, one recovers from $V(c)$ the expression given in \cite{weiss}
for the effective potential, $V(c) \rightarrow (4\pi^{2} /3) c^2 (1-c)^2 $.
In this same limit, the term in the large parentheses in (\ref{ergeb})
entering $W(c)$ is dominated by its first summand. Figure \ref{fig} displays
the effective potential $V_{eff} $ as a function of $c$ for various large 
$\beta \Lambda $ at a suitably small coupling $g$,
which is the regime one is ultimately interested in when 
taking the cutoff $\Lambda $ to infinity. The value $c=1/2$ corresponds
to vanishing Polyakov loop order parameter. One observes a
second-order transition between a regime of larger $\beta \Lambda $,
where the minimum of the effective potential is at $c=1/2$, and a regime
of smaller $\beta \Lambda $, where the order parameter develops a
nonvanishing expectation value due to spontaneous breaking of the center
symmetry in the double-well potential. The precise value of $\beta \Lambda $
at which the transition takes place rises as the coupling constant
$g$ falls. Some care should be taken in interpreting the role of $g$.
It is possible to adjust $g$ as the cutoff $\Lambda $ is taken to
infinity such as to obtain finite results for observables (more specific
remarks on this follow further below). However, one should not expect
the renormalization to mirror the behavior familiar e.g. from perturbation
theory; note that the loop expansion carried out for the vector
fields cannot be equated with the usual perturbative expansion, since
the field $c$ entering the Polyakov loop is of order $1/g$.
Furthermore, the renormalization of $g$ in general also depends on
the inverse temperature $\beta $. The authors have not explored the 
possibility of establishing a renormalization group type of equation
relating renormalized coupling constants $g$ at different $\beta $.
At this stage, the effective theory can be used as follows: Fit the 
temperature dependent running coupling $g$ such as to reproduce the
temperature dependence of a physical quantity, e.g. the string tension
in the confining phase, always in the physical limit
$\Lambda \rightarrow \infty $. Then other quantities can be predicted.
The authors have tested the feasibility of this scheme by deriving
the string tension from the Polyakov loop correlator. Its negative
logarithm, $-(1/\beta ) \ln \langle L(\vec{R} ) L(0) \rangle $, can be
interpreted \cite{ben} as the free energy associated with a pair of
static sources separated by the vector $\vec{R} $. On the other hand, for
small oscillations of the field $c$ around $c=1/2$ in the confined
phase one has the estimate $\langle L(\vec{R} ) L(0) \rangle =
\langle u(\vec{R} ) u(0) \rangle \pi^{2} \beta /W(c=1/2)$.
Furthermore, in this approximation one can replace the effective
action for $u$ by $\int d^3 x \, (\partial_{i} u \partial_{i} u
+\beta^{2} \sigma^{2} u^2 )$ with
$\beta^{2} \sigma^{2} = 
(\partial^{2} V_{eff} / \partial c^2 )/(2W) |_{c=1/2} $.
Then the propagator $\langle u(\vec{R} ) u(0) \rangle $ takes the 
simple Yukawa form $\langle u(\vec{R} ) u(0) \rangle =
e^{-\beta \sigma R} /(16\pi R)$, so that $\sigma $ can be directly
identified with the string tension. In order to reproduce a finite
string tension as the temperature approaches zero and a finite
deconfinement temperature, it turns out that one should scale
$g(\beta /\beta_{cr} \rightarrow 1) \sim 1/(\beta_{cr} \Lambda )^2 $ and
$g(\beta /\beta_{cr} \rightarrow \infty ) \sim 1/(\beta_{cr} \Lambda ) $
where $\beta_{cr} $ is the inverse deconfinement temperature.

To conclude, a quite drastic approximation on the dynamics of the vector
fields has been shown to nevertheless reproduce the confining and
deconfining phases of Yang-Mills theory. It appears that the confining
phase can be simply understood as a consequence of the metric
properties of the gauge group, encoded in the centrifugal barriers
entering the kinetic terms, cf. eq. (\ref{eform}). The detailed dynamics
of the vector fields $A_i $ were largely ignored in the present
approximation. The picture obtained here for the emergence of the
confined phase stands in contrast to the Z(2) bubble mechanism
advocated on the basis of previous work \cite{weiss}. In the
low-temperature regime, there is no need for a Z(2) domain structure
to restore center symmetry. However, Z(2) bubbles may be relevant
for the existence of a superheated confining phase above what one
would usually consider the deconfinement temperature. The confinement
mechanism in this phase would be akin to a localization phenomenon.
A further comment concerns the formulation of high-temperature
perturbation theory. There, one usually assumes small fluctuations
of the field $a_0 $ around $a_0 =0$. Clearly, this is not correct
even in the deconfined phase, since the value $a_0 =0$ is always
protected by centrifugal barriers; one should therefore expand
around a nontrivial value for $a_0 $ as given by the effective
potential (\ref{veffu}). It is tempting to speculate \cite{lenzneu}
that this is the origin of the infrared problems initially reported by
Linde \cite{lind}. Work is in progress to calculate perturbative
corrections to the effective potential, to extend the treatment to
SU(3) color, and to determine how other observables, such as the
gluon condensate, behave under the present approximation scheme.
Also a more detailed treatment of the renormalization is under study.

{\bf Acknowledgments.}
The authors are grateful to L.Gamberg, H.Gies and K.Langfeld
for useful discussions and also acknowledge the latter for
communicating his lattice results. R.Alkofer is thanked for a careful
reading of the manuscript. This work was
supported by Deutsche Forschungsgemeinschaft under DFG Re 856 / 1-3.

\end{document}